\documentclass[12pt,english,floatfix,nofootinbib,superscriptaddress,aps,prd,preprint]{revtex4}
\usepackage[utf8]{inputenc}
\usepackage{float}
\usepackage{array}
\usepackage{lipsum}
\usepackage{graphicx}
\usepackage{amsmath,amsthm,amsfonts,amssymb}
\usepackage{tikz}
\usepackage{multirow}

\newcommand{\be}{\begin{equation}}
\newcommand{\ee}{\end{equation}}
\newcommand{\Be}{\begin{eqnarray}}
\newcommand{\Ee}{\end{eqnarray}}

\newcommand{\mincir}{\raise
-3.truept\hbox{\rlap{\hbox{$\sim$}}\raise4.truept\hbox{$<$}\ }}
\newcommand{\magcir}{\raise
-3.truept\hbox{\rlap{\hbox{$\sim$}}\raise4.truept\hbox{$>$}\ }}

\newcolumntype{Y}{>{\centering\arraybackslash}X}
\providecommand{\U}[1]

 \usetikzlibrary{quotes,angles}
 \usetikzlibrary{arrows} 
\usetikzlibrary{decorations.markings}
\usepackage{graphicx}
\usepackage[english]{babel}
\usepackage{color}
\usepackage{subfig}
\usepackage{caption}
\usepackage{tensor}
\usepackage{esint}
\usepackage[dvips]{epsfig}
\usepackage[dvips]{graphicx}
\usepackage{float}
\usepackage{units}
\usepackage{textcomp}
\usepackage{mathrsfs}
\usepackage{amsmath}
\usepackage[makeroom]{cancel}
\usepackage{amssymb}
\usepackage{amsbsy}
\usepackage{amsfonts}
\usepackage{amssymb,mathrsfs,xcolor}
\usepackage{esint}
\usepackage{array}
\usepackage{graphicx}

\usepackage{hyperref}
\hypersetup{
    colorlinks,
    citecolor=blue,
    filecolor=green,
    linkcolor=purple,
    urlcolor=red,
}

\usepackage{slashed}

\newcommand{\ie}{\begin{equation}}
\newcommand{\fe}{\end{equation}}
\newcommand{\se}{\begin{eqnarray}}
\newcommand{\ff}{\end{eqnarray}}

\usepackage{hyperref}
\hypersetup{colorlinks,breaklinks,
			citecolor=[rgb]{0,0.0,1.0},
            urlcolor=[rgb]{0.0,0.0,1.0},
            linkcolor=[rgb]{0,0.5,0.9}}

\begin{document}

\title{Antisymmetric tensor influence on charged black hole lensing phenomena and time delay}

\author{A. A. Ara\'{u}jo Filho}
\email{dilto@fisica.ufc.br}
\affiliation{Departamento de Física, Universidade Federal da Paraíba, Caixa Postal 5008, 58051--970, João Pessoa, Paraíba,  Brazil.}


\date{\today}

\begin{abstract}

In this work, we investigate the gravitational lensing of a charged black hole (spherically symmetric) within the context of Lorentz violation triggered by an antisymmetric tensor field. Our calculations consider two different scenarios: the weak and strong deflection limits. For the weak deflection limit, we employ the \textit{Gauss--Bonnet} theorem to obtain our results. In the strong deflection limit, we utilize the \textit{Tsukamoto} methodology, which provides measurable outcomes such as relativistic image positions and magnifications. Applying the latter methodology, we analyze the gravitational lensing by Sagittarius $A^{*}$ and derive the related observables, which are expressed as functions of the Lorentz violation parameter. Finally, the time delay is addressed as well.


\end{abstract}


\maketitle


\section{Introduction}
\label{sec:intro}

Lorentz symmetry, a foundational principle in contemporary physics, asserts that physical laws remain invariant across all inertial frames. Nevertheless, deviations from Lorentz symmetry have been theorized under specific energy conditions within diverse theoretical frameworks. These involve very special relativity \cite{x08}, Einstein--aether theory \cite{x05}, non--commutative field theory \cite{x04}, string theory \cite{x01}, Horava--Lifshitz gravity \cite{x03},  \(f(T)\) gravity \cite{x07}, massive gravity \cite{x06},  loop quantum gravity \cite{x02} and other theoretical models. The violation of Lorentz symmetry presents itself in two distinct manners: explicit and spontaneous \cite{bluhm2006overview}. Explicit violation occurs when the Lagrangian density lacks Lorentz invariance, resulting in the formulation of distinct physical laws in specific reference frames. Conversely, spontaneous violation occurs when the Lagrangian density preserves Lorentz invariance, but the ground state of a physical system does not exhibit Lorentz symmetry \cite{bluhm2008spontaneous}.

The investigation into spontaneous Lorentz symmetry breaking \cite{b9,b10,b11,b12,b13,KhodadiPoDU2023} is grounded in the framework of the Standard Model Extension. In this theoretical framework, basic field theories are encapsulated by bumblebee models \cite{b1,b10,b11,b12,b13, KhodadiEPJC2023,araujo2024exact,Filho:2022yrk,KhodadiEPJC20232,KhodadiPRD2022,CapozzielloJCAP2023}. These models feature a vector field, termed the bumblebee field, which acquires a non--zero vacuum expectation value (VEV). This property establishes a preferred direction, leading to the breakdown of local Lorentz invariance for particles and consequently resulting in significant implications, including effects on thermodynamic properties \cite{aa2021lorentz,araujo2021thermodynamic,aa2022particles,reis2021thermal,araujo2021higher,araujo2022thermal,paperrainbow,anacleto2018lorentz,araujo2023thermodynamics}.

Beyond the conventional vector field theories, an alternative approach to investigating Lorentz symmetry breaking (LSB) involves the study of a rank--two antisymmetric tensor field known as the Kalb--Ramond field \cite{042, assunccao2019dynamical, maluf2019antisymmetric,AraujoFilho:2024ctw}. This field is inherently present in the spectrum of bosonic string theory \cite{043}. When it is non--minimally coupled to gravity and acquires a non--zero vacuum expectation value, spontaneous breaking of Lorentz symmetry is induced. An exact solution for a static, spherically symmetric configuration within this framework has been presented by Ref. \cite{044}. Subsequently, an extensive investigation into the dynamics of both massive and massless particles near such static spherical Kalb--Ramond black holes has been conducted by Ref. \cite{045}. Furthermore, the gravitational deflection of light and the shadows cast by rotating black holes within this theoretical framework have been explored by Ref. \cite{046}. Additionally, recent literature has addressed the Generalized Uncertainty Principle (GUP) corrections, gravitational parity violations involving antisymmetric tensors, and the cosmological implications of Kalb--Ramond--like particles \cite{baruah2023quasinormal, manton2024kalb, capanelli2023cosmological}.

Recent developments, highlighted by the detection of gravitational waves by collaborations such as LIGO--Virgo \cite{016,017,018}, have significantly expanded the scope of cosmological research. Gravitational waves are now essential tools for exploring the universe, including the study of gravitational lensing within the weak field approximation \cite{019,020}. Traditionally, research on gravitational lensing has focused on light traveling over vast distances from gravitational sources, such as in Schwarzschild spacetime \cite{021}, and later extended to encompass general spherically symmetric and static spacetimes \cite{022}. Nevertheless, whenerver there exits regions with strong gravitational fields, particularly near black holes, the angular deviation of light is significantly magnified, as one should expect.

Recent observations made by the Event Horizon Telescope of a supermassive black hole situated at the heart of the M87 galaxy have generated profound scientific interest \cite{023,024,025,026,027,028,029}. The foundational work by Virbhadra and Ellis introduced a concise lens equation tailored for supermassive black holes within an asymptotically flat background \cite{030,031}, revealing multiple symmetrically distributed images around the optical axis due to strong gravitational effects. Subsequent advancements spearheaded by Fritelli et al. \cite{032}, Bozza et al. \cite{033}, and Tsukamoto \cite{035} have refined the analytical frameworks for investigating strong field gravitational lensing. These studies have explored light deflection across various scenarios \cite{virbhadra2002gravitational,virbhadra1998role,virbhadra2000schwarzschild,grespan2023strong,cunha2018shadows,oguri2019strong,metcalf2019strong,bisnovatyi2017gravitational,ezquiaga2021phase}, including exotic structures such as wormholes \cite{ovgun2019exact,38.1,38.2,38.3,38.4,38.5}, rotating solutions \cite{hsieh2021strong,hsieh2021gravitational,jusufi2018gravitational,37.1,37.2,37.3,37.4,37.5,37.6}, alternative theories of gravity \cite{chakraborty2017strong,t25:1,40,nascimento2024gravitational}, and Reissner--Nordström spacetime \cite{036,036.1,036.2}, and others \cite{zhang2024strong,tsukamoto2023gravitational}. Also, gravitational distortion has also been addressed \cite{virbhadra2024conservation,virbhadra2022distortions}

In the past decades, a significant surge of interest in the exploration of gravitational waves and their spectra, as highlighted in recent scientific literature \cite{bombacigno2023landau,aa2023analysis,boudet2022quasinormal,aa2024implications,hassanabadi2023gravitational,amarilo2023gravitational}. This increased focus is rooted in substantial advancements in the technology used to detect gravitational waves, exemplified notably by the VIRGO and LIGO detectors. These sophisticated instruments have played a pivotal role in yielding profound insights into the intriguing realm of black hole physics \cite{abbott2016gw150914,abramovici1992ligo,grishchuk2001gravitational,vagnozzi2022horizon}. Yang et al. has introduced innovative exact solutions for static and spherically symmetric spacetimes, both in the presence and absence of a cosmological constant, within the context of a non--zero vacuum expectation value background of the Kalb--Ramond field \cite{yang2023static}. Furthermore, their study extends to proposing a model for charged black holes within this theoretical framework \cite{duan2023electrically}.

Very recently in the literature, authors conducted calculations on the gravitational lensing of an uncharged black hole (originally proposed in \cite{yang2023static}), exploring both weak and strong deflection limits \cite{junior2024gravitational}. Building upon there works, we expand this investigation by considering a charged black hole instead \cite{duan2023electrically}. To accomplish this, we follow a similar approach. Essentially, we compute the gravitational lensing effects in both regimes: the weak deflection limit and the strong deflection limit. In addition, the calculation of the time delay is also presented.

\section{The general setup}

Fundamentally, in this brief section, we present both black holes that we shall take into account to develop our calculations. Initially, we consider a spherically symmetric (charged) solution within the context of Lorentz violation triggered by an antisymmetric tensor field, without a cosmological constant \cite{duan2023electrically}
\ie
\label{metrictsuka}
\mathrm{d}s^{2} = - \left( \frac{1}{1-l} - \frac{2M}{r} + \frac{Q^{2}}{(1-l)^{2}r^{2}}    \right) \mathrm{d}t^{2} + \frac{\mathrm{d}r^{2}}{\left( \frac{1}{1-l} - \frac{2M}{r} + \frac{Q^{2}}{(1-l)^{2}r^{2}}    \right)} + r^{2} \mathrm{d} \theta^{2} + r^{2}\sin^{2}\theta \mathrm{d}\varphi^{2}.
\fe
It is worth mentioning that recent studies have addressed the implications of charged and uncharged metrics by considering, for instance, thermodynamics, phase transitions, geodesics, shadows, scattering effects, and \textit{quasinormal} modes \cite{yang2023static,araujo2024exploring,heidari2024impact}.


\section{Gravitational lensing via weak deflection limit}


In this section, we conduct an examination of the \textit{Gauss--Bonnet} theorem and proceed to calculate the weak deflection angle of the black hole \cite{Gibbons:2008rj}. Adopting this framework, one can describe the trajectories of massless particles—those governed by the condition $\mathrm{d}s^2 = 0$—through the following expression, as detailed in Refs. \cite{araujo2024effects, Heidari:2025iiv, heidari2024absorption}:
\ie
\mathrm{d}t^2=\gamma_{ij}\mathrm{d}x^i \mathrm{d}x^j = -\frac{g_{rr}(r)}{g_{tt}(r)}\mathrm{d}r^2  -\frac{\Bar{g}_{\varphi\varphi}(r)}{g_{tt}(r)}\mathrm{d}\varphi^2.
\fe

In this context, the indices $i$ and $j$ run over the spatial coordinates $1$ to $3$, and the quantity $\gamma_{ij}$ denotes the components of the optical metric. For convenience, the angular component evaluated on the equatorial plane is defined as $\Bar{g}_{\varphi\varphi}(r) \equiv g_{\varphi\varphi}(r, \theta = \pi/2)$. Moreover, the Gaussian curvature associated with the optical geometry is expressed as \cite{qiao2024existence}:
\ie
\label{dffdsf}
\mathcal{K}(r,l,Q) = \frac{R}{2} =  \frac{g_{tt}(r)}{\sqrt{g_{rr}(r) \,  \Bar{g}_{\varphi\varphi}(r)}}  \frac{\partial}{\partial r} \left[  \frac{g_{tt}(r)}{2 \sqrt{g_{rr}(r) \, \Bar{g}_{\varphi\varphi}(r) }}   \frac{\partial}{\partial r} \left(   \frac{\Bar{g}_{\varphi\varphi}(r)}{g_{tt}(r)}    \right)    \right].
\fe
Here, $R$ denotes the two--dimensional Ricci scalar associated with the optical geometry. In the regime where the parameter $l$ is small, the curvature can be approximated by the following expression \cite{heidari2024impact,al-Badawi:2024pdx}:
\ie
\begin{split}
\label{gaussiancurvature}
& \mathcal{K}(r,l,Q) = +\frac{3 Q^2}{(1-l)^3 r^4}+\frac{3 M^2}{r^4} -\frac{6 M Q^2}{(1-l)^2 r^5}-\frac{2 M}{(1-l) r^3}+\frac{2 Q^4}{(1-l)^4 r^6}.
\end{split}
\fe
The area of the surface on the equatorial plane is given by \cite{qiao2024existence,Heidari:2025iiv}:
\begin{equation}
\mathrm{d}S = \sqrt{\gamma} \, \mathrm{d} r \mathrm{d}\varphi = \sqrt{\frac{g_{rr}(r)}{g_{tt} (r)}  \frac{g_{\varphi\varphi}(r)}{g_{tt}(r)} } \, \mathrm{d} r \mathrm{d}\varphi. 
\end{equation}

With all these preliminaries, the deflection angle within the context of weak deflection limit reads
\ie
\begin{split}
\label{lensing1}
\alpha (b,l,Q) & = -\int\int\mathcal{K}\mathrm{d}S=-\int^{\pi}_0\int^{\infty}_{ r = \frac{b}{\sin\phi}}\mathcal{K}\mathrm{d}S \\
\simeq &  \frac{4 M}{b}  +\frac{3 \pi  M^2}{4 b^2} -  \frac{2 l M}{b}  - \frac{9 \pi  l M^2}{8 b^2}  -\frac{9 \pi  l Q^2}{8 b^2}-\frac{3 \pi  Q^2}{4 b^2}  \\
& -   \frac{8 M Q^2}{3 b^3} -\frac{4 l M Q^2}{3 b^3} + \frac{45 \pi  l M^2 Q^2}{64 b^4} - \frac{45 \pi  M^2 Q^2}{32 b^4} .
\end{split}
\fe

\begin{figure}
    \centering
     \includegraphics[scale=0.51]{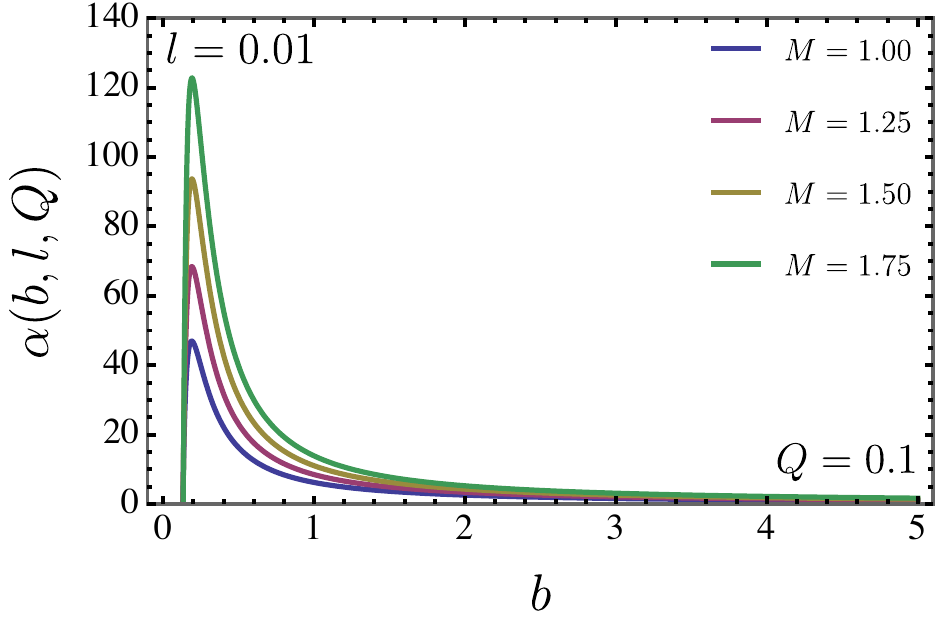}
    \includegraphics[scale=0.51]{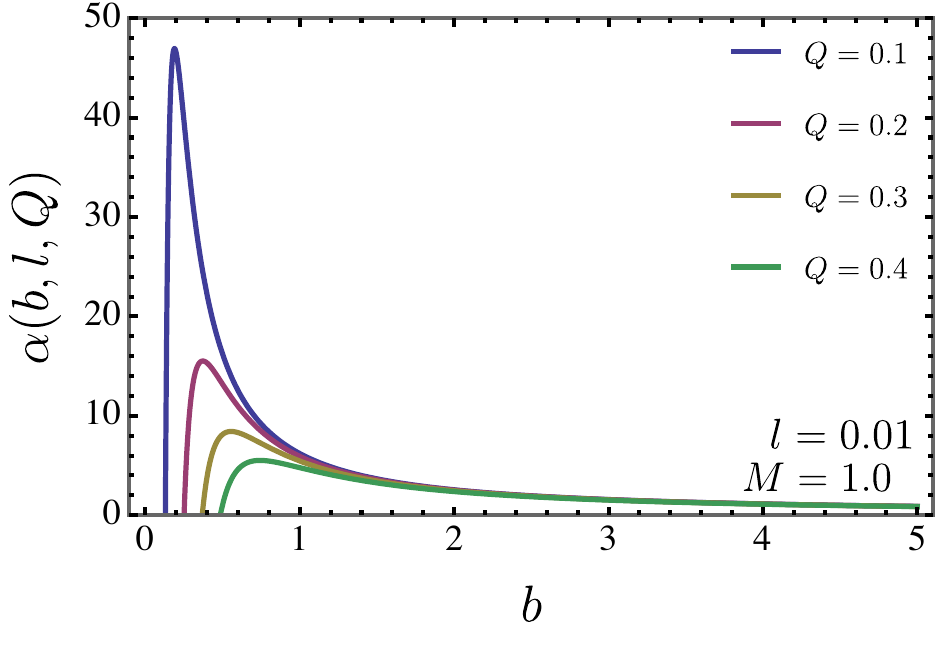}
     \includegraphics[scale=0.51]{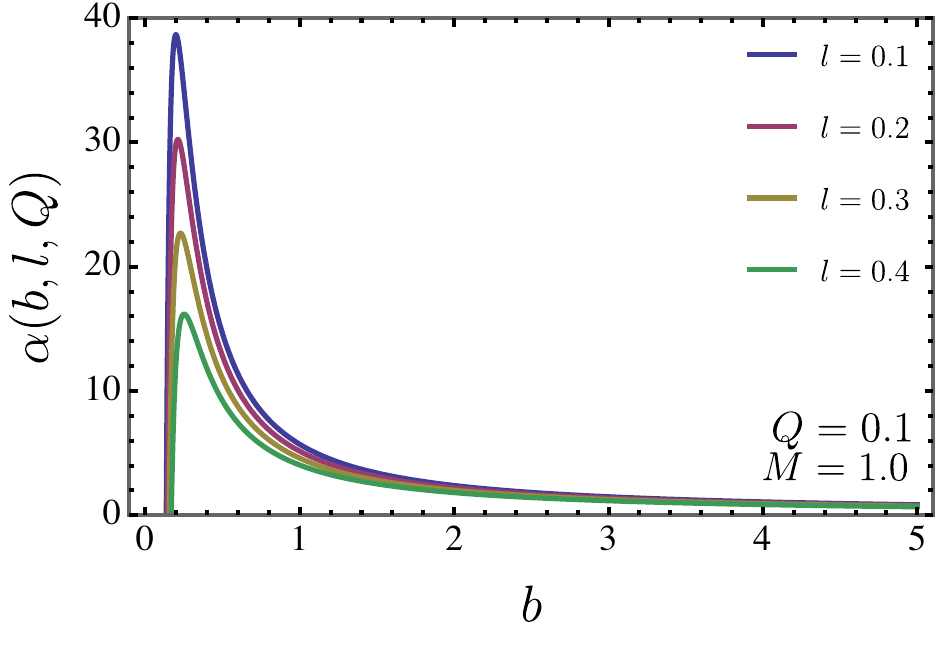}
    \caption{The deflection angle as a function of $b$ for different values of $l$, $M$ and $Q$.}
    \label{asdasd}
\end{figure}

To obtain the above expression, we considered the same limit as in Ref. \cite{Gibbons:2008rj}, specifically $b \gg 2M$. Note that the first two terms correspond to the Schwarzschild case calculation (up to the second order in $M$). The terms containing $l$ are introduced by Lorentz violation, while the terms with $Q$ (but without $l$) correspond to those found in the Reissner--Nordström black hole. To further support the interpretation of Eq. (\ref{lensing1}), we present Fig. \ref{asdasd}. In the top--left panel, we observe that increasing the mass results in a greater magnitude of the weak deflection angle. Conversely, the top--right panel shows that an increase in charge leads to a decrease in the magnitude of $\alpha(b, l, Q)$. Additionally, the bottom panel demonstrates that as $l$ increases, the magnitude of $\alpha(b, l, Q)$ also decreases.

An additional remark worth considering is that it is not mandatory to adopt $b/\sin \phi$ as the lower integration limit in Eq. (\ref{lensing1}). Recent studies, for instance, have extended the analysis to include higher--order terms \cite{Jha:2024ltc,AraujoFilho:2024mvz}
\ie
u = \frac{1}{r} = \frac{\sin\phi}{b} + \frac{M(1-\cos\phi)^2}{b^2}-\frac{M^2(60\phi\,
\cos\phi+3\sin3\phi-5\sin\phi)}{16b^3}.
\fe

\section{Gravitational lensing via the strong deflection limit}

The approach employed in this part of the analysis aims to derive the expression for the bending angle of light in the regime of strong gravitational deflection \cite{nascimento2024gravitational, heidari2024absorption}. Among the established frameworks available, we adopt the procedure initially developed by Tsukamoto \cite{tsukamoto2017deflection}, which is particularly suited to probe photon trajectories in regions where the gravitational field is extremely intense. The spacetime geometry under consideration is defined by a general line element of the form:
\ie
\label{ssst}
\mathrm{d}s^{2} = - \mathcal{A}(r) \mathrm{d}t^{2} + \mathcal{B}(r) \mathrm{d}r^{2} + \mathcal{C}(r)(\mathrm{d}\theta^2 + \sin^{2}\theta\mathrm{d}\phi^2).
\fe

In order to implement the formalism established by Tsukamoto \cite{tsukamoto2017deflection}, one must ensure that the underlying spacetime geometry adheres to the condition of asymptotic flatness in principle. This imposes specific constraints on the metric functions: as $r$ approaches infinity, the temporal and radial components must converge to unity, i.e., $\mathcal{A}(r) \to 1$ and $\mathcal{B}(r) \to 1$, while the angular part should asymptotically approach $\mathcal{C}(r) \to r^2$. The static and spherically symmetric nature of the geometry implies the existence of conserved quantities associated with the Killing vectors $\partial_t$ and $\partial_\phi$, which correspond to energy and angular momentum, respectively.

To advance toward the derivation of the deflection angle in the strong deflection regime, we begin by introducing an auxiliary function $\mathcal{D}(r)$, which simplifies the mathematical structure of the integral formulation used in the analysis
\ie
\mathcal{D}(r) \equiv \frac{\mathcal{C}^{\prime}(r)}{\mathcal{C}(r)} - \frac{\mathcal{A}^{\prime}(r)}{\mathcal{A}(r)}.
\fe
Derivatives with respect to the radial coordinate are denoted by the prime ($\prime$) symbol. It is presumed that the equation $\mathcal{D}(r) = 0$ admits at least one positive root. Among these, the photon sphere is identified by selecting the greatest such root, which we label as $r_{ph}$. For the analysis to remain valid in the region extending from the photon sphere outward, the metric functions $\mathcal{A}(r)$, $\mathcal{B}(r)$, and $\mathcal{C}(r)$ are required to be well--defined, strictly positive, and finite for all $r \geq r_{ph}$.

The symmetries of the spacetime, expressed through the Killing vectors $\partial_t$ and $\partial_\phi$, lead to the conservation of two physical quantities along the null geodesics: the energy $E$ and the angular momentum $L$. These are given by $E = \mathcal{A}(r)\dot{t}$ and $L = \mathcal{C}(r)\dot{\phi}$, respectively. Assuming that neither $E$ nor $L$ vanishes, one can define the impact parameter $b$, which characterizes the trajectory of the photon, as follows:
\ie
b \equiv \frac{L}{E} = \frac{\mathcal{C}(r)\Dot{\phi}}{\mathcal{A}(r)\Dot{t}}.
\fe

Taking advantage of the axial symmetry of the spacetime, the analysis can be confined to the equatorial plane by setting $\theta = \pi/2$, without any loss of generality. Under this condition, the geodesic equation simplifies, allowing the radial component of the motion to be expressed as:
\ie
\Dot{r}^{2} = V(r).
\fe
By defining the function $R(r)$ as $R(r) \equiv \mathcal{C}(r)/[\mathcal{A}(r)b^{2}] - 1$, one can express the radial equation in terms of an effective potential, $V(r)$, given by $V(r) = L^2 R(r)/[\mathcal{B}(r)\mathcal{C}(r)]$. This formulation mirrors the classical motion of a particle with unit mass under the influence of a potential $V(r)$. The photon’s motion is therefore constrained to regions where $V(r) \geq 0$. The asymptotic behavior of the spacetime ensures that, as $r \to \infty$, the potential tends toward a positive constant: $V(r) \to E^2 > 0$, indicating that photons can exist at arbitrarily large distances.

We concentrate on the lensing configuration where a photon, emitted from a distant source, travels toward the compact object, reaches a minimal radial coordinate $r_0$, and then escapes back to spatial infinity. For such a deflection process to take place rather than being trapped in a circular orbit, it is required that $r_0 > r_{ph}$. The turning point of the trajectory, $r = r_0$, is characterized as the outermost positive root of $R(r) = 0$, at which the effective potential satisfies $V(r_0) = 0$. Provided that the functions $\mathcal{B}(r)$ and $\mathcal{C}(r)$ are regular at this point, one can infer from the definition of $R(r)$ the following condition at the point of closest approach:
\ie
\mathcal{A}_{0}\Dot{t}^{2}_{0} = \mathcal{C}_{0}\Dot{\phi}^{2}_{0}.
\fe

It is worth mentioning that any quantity carrying the subscript "$0$" is understood to be evaluated at the radial distance $r = r_0$, which represents the point of closest approach of the photon to the lensing object. Since we are interested in the trajectory of a single light ray, we take the impact parameter $b$ to be positive—an assumption that does not affect the generality of the treatment. Given that $b$ is a positive quantity along the path of the photon, it can be determined directly at $r_0$ by the following relation:
\ie
b(r_{0}) = \frac{L}{E} = \frac{\mathcal{C}_{0}\Dot{\phi}_{0}}{\mathcal{A}_{0}\Dot{t}_{0}} = \sqrt{\frac{\mathcal{C}_{0}}{\mathcal{A}_{0}}}.
\fe

It is also worth noting that the function $R(r)$ can alternatively be written in the equivalent form:
\ie
R(r)= \frac{\mathcal{A}_{0}\mathcal{C}}{\mathcal{A}\mathcal{C}_{0}} - 1.
\fe

Following the methodology introduced in Ref. \cite{hasse2002gravitational}, one can establish a criterion that fully characterizes the presence of circular photon orbits. This condition is not only required but also guarantees the existence of such trajectories. Based on this framework, the corresponding equation governing the path of light takes the form:
\ie
\frac{\mathcal{B} \mathcal{C} \Dot{r}^{2}}{E^{2}} + b^{2} = \frac{\mathcal{C}}{\mathcal{A}},
\fe
so that
\ie
\ddot{r} + \frac{1}{2}\left( \frac{\mathcal{B}^{\prime}}{\mathcal{B}} + \frac{\mathcal{C}^{\prime}}{\mathcal{C}} \Dot{r}^{2} \right) = \frac{E^{2}\mathcal{D}}{\mathcal{A}\mathcal{B}}. 
\fe

In the region where $r \geq r_{ph}$, the metric functions $\mathcal{A}(r)$, $\mathcal{B}(r)$, and $\mathcal{C}(r)$ are assumed to be regular, strictly positive, and finite. Given a positive energy $E$, the condition $\mathcal{D}(r) = 0$ serves as the fundamental requirement for the existence of a circular photon orbit. At the radius of the photon sphere, $r = r_{ph}$, we find that the derivative of $R(r)$ vanishes, expressed as $R'_{ph} = \mathcal{D}_{ph} \mathcal{C}_{ph} \mathcal{A}_{ph}/b^2 = 0$. Here, the subscript “$ph$” indicates evaluation at the photon sphere.

We now introduce the critical value of the impact parameter, denoted as $b_c$, which is 
\ie
b_{c}(r_{ph}) \equiv \lim_{r_{0} \to r_{ph}} \sqrt{\frac{\mathcal{C}_{0}}{\mathcal{A}_{0}}}.
\fe

From this point onward, the scenario under consideration will be classified as the strong deflection regime. To proceed further, we compute the radial derivative of the effective potential $V(r)$, which yields:
\ie
V^{\prime}(r) = \frac{L^{2}}{\mathcal{B}\mathcal{C}} \left[ R^{\prime} + \left( \frac{\mathcal{C}^{\prime}}{\mathcal{C}} - \frac{\mathcal{B}^{\prime}}{\mathcal{B}}   \right)   R  \right].
\fe

In the regime of strong gravitational deflection, as the point of closest approach $r_0$ approaches the photon sphere radius $r_{ph}$, both the effective potential $V(r_0)$ and its radial derivative $V'(r_0)$ simultaneously vanish. This limiting behavior reshapes the equation governing the light's path into the simplified expression:
\ie
\left(  \frac{\mathrm{d}r}{\mathrm{d}\phi}     \right)^{2} = \frac{R(r)\mathcal{C}(r)}{\mathcal{B}(r)}.
\fe
Accordingly, the quantity known as the deflection angle, denoted by $\alpha(r_0)$, takes the following integral form in terms of the closest approach distance $r_0$:
\ie
\alpha(r_{0}) = I(r_{0}) - \pi,
\fe
with the function $I(r_0)$, which encodes the integral contribution to the deflection angle, is defined as follows:
\ie
I(r_{0}) \equiv 2 \int^{\infty}_{r_{0}} \frac{\mathrm{d}r}{\sqrt{\frac{R(r)\mathcal{C}(r)}{\mathcal{B}(r)}}}.
\fe

As a first step in the analysis, attention must be turned to the integral expression governing the deflection angle. However, as emphasized in Tsukamoto’s investigation \cite{tsukamoto2017deflection}, evaluating this integral analytically is far from straightforward due to the divergence that arises near the photon sphere. To facilitate the computation, we introduce the following auxiliary function, as also suggested in Ref. \cite{tsukamoto2017deflection}:
\ie
z \equiv 1 - \frac{r_{0}}{r}.
\fe
Here, with this new definition in place, the integral expression can be reformulated in the following manner:
\ie
I(r_{0}) = \int^{1}_{0} f(z,r_{0}) \mathrm{d}z,
\fe
where 
\ie
f(z,z_{0}) \equiv \frac{2r_{0}}{\sqrt{G(z,r_{0})}}, \,\,\,\,\,\,\,\, \text{and} \,\,\,\,\,\,\,\,  G(z,r_{0}) \equiv R \frac{\mathcal{C}}{\mathcal{B}}(1-z)^{4}.
\fe

Expressed in terms of the variable $z$, the function $R(r)$ can be rewritten as:
\ie
R(r) = \mathcal{D}_{0}r_{0} z + \left[ \frac{r_{0}}{2}\left( \frac{\mathcal{C}^{\prime\prime}_{0}}{\mathcal{C}_{0}} - \frac{\mathcal{A}_{0}^{\prime\prime}}{\mathcal{A}_{0}}  \right) + \left( 1 - \frac{\mathcal{A}_{0}^{\prime}r_{0}}{\mathcal{A}_{0}}  \right) D_{0}  \right] r_{0} z^{2} + \mathcal{O}(z^{3})+ ...    \,\,\,\,.
\fe

We now proceed by expanding the function $G(z, r_0)$ in a power series around $z = 0$. This yields:
\ie
G(z,r_{0}) = \sum^{\infty}_{n=1} c_{n}(r_{0})z^{n},
\fe
in which the quantities $c_1(r)$ and $c_2(r)$ are expressed as
\ie
c_{1}(r_{0}) = \frac{\mathcal{C}_{0}\mathcal{D}_{0}r_{0}}{\mathcal{B}_{0}},
\fe
and
\ie
c_{2}(r_{0}) = \frac{\mathcal{C}_{0}r_{0}}{\mathcal{B}_{0}} \left\{ \mathcal{D}_{0} \left[ \left( \mathcal{D}_{0} - \frac{\mathcal{B}^{\prime}_{0}}{\mathcal{B}_{0}}  \right)r_{0} -3       \right] + \frac{r_{0}}{2} \left(  \frac{\mathcal{C}^{\prime\prime}_{0}}{\mathcal{C}_{0}} - \frac{\mathcal{A}^{\prime\prime}_{0}}{\mathcal{A}_{0}}  \right)                 \right\}.
\fe

Additionally, in the regime of strong gravitational deflection, one finds that
\ie
c_{1}(r_{ph}) = 0, \,\,\,\,\,\, \text{and} \,\,\,\,\,\, c_{2}(r_{ph}) =  \frac{\mathcal{C}_{ph}r^{2}_{ph}}{2 \mathcal{B}_{ph}}\mathcal{D}^{\prime}_{ph}, \,\,\,\,\,\,\, \text{with} \,\,\,\,\, \mathcal{D}^{\prime}_{ph} = \frac{\mathcal{C}^{\prime\prime}}{\mathcal{C}_{ph}} - \frac{\mathcal{A}^{\prime\prime}}{\mathcal{A}_{ph}}.
\fe
In this context, $G(z, r_0)$ is written in a more compact form, as follows:
\ie
G_{ph}(z) = c_{2}(r_{ph})z^{2} + \mathcal{O}(z^{3}).
\fe

The divergence in the function $f(z, r_0)$ becomes evident through its leading behavior, which scales as $1/z$ near $z = 0$. This singularity gives rise to a logarithmic divergence in the integral $I(r_0)$ as the closest approach radius $r_0$ approaches the photon sphere radius $r_{ph}$. To handle this issue analytically, the integral $I(r_0)$ is decomposed into two separate contributions: a singular term, denoted by $I_D(r_0)$, which takes into account the divergent structure, and a finite remainder, referred to as $I_R(r_0)$. Accordingly, the divergent component takes the form:
\ie
I_{D}(r_{0}) \equiv \int^{1}_{0} f_{D}(z,r_{0}) \mathrm{d}z, \,\,\,\,\,\,\, \text{with} \,\,\,\,\,\,f_{D}(z,r_{0}) \equiv \frac{2 r_{0}}{\sqrt{c_{1}(r_{0})z + c_{2}(r_{0})z^{2}}}.
\fe
Upon performing the integration, the result becomes
\ie
I_{D} (r_{0}) = \frac{4 r_{0}}{\sqrt{c_{2}(r_{0})}} \ln \left[  \frac{\sqrt{c_{2}(r_{0})} + \sqrt{c_{1}(r_{0}) + c_{2}(r_{0})     }  }{\sqrt{c_{1}(r_{0})}}  \right].
\fe

Expanding $c_1(r_0)$ and $b(r_0)$ in a series around the point $r_0 = r_{ph}$
\ie
c_{1}(r_{0}) = \frac{\mathcal{C}_{ph}r_{ph}\mathcal{D}^{\prime}_{ph}}{\mathcal{B}_{ph}} (r_{0}-r_{ph}) + \mathcal{O}((r_{0}-r_{ph})^{2}),
\fe
and
\ie
b(r_{0}) = b_{c}(r_{ph}) + \frac{1}{4} \sqrt{\frac{\mathcal{C}_{ph}}{\mathcal{A}_{ph}}}\mathcal{D}^{\prime}_{ph}(r_{0}-r_{ph})^{2} + \mathcal{O}((r_{0}-r_{ph})^{3}),
\fe
which yields the following expression in the strong deflection regime
\ie
\lim_{r_{0} \to r_{ph}} c_{1}(r_{0})  =  \lim_{b \to b_{c}} \frac{2 \mathcal{C}_{ph} r_{ph} \sqrt{\mathcal{D}^{\prime}}}{B_{ph}} \left(  \frac{b}{b_{c}} -1  \right)^{1/2}.
\fe

In this way, we write $I_{D}(b)$ as being
\ie
I_{D}(b) = - \frac{r_{ph}}{\sqrt{c_{2}(r_{ph})}} \ln\left[ \frac{b}{b_{c}} - 1 \right] + \frac{r_{ph}}{\sqrt{c_{2}(r_{ph})}}\ln \left[ r^{2}\mathcal{D}^{\prime}_{ph}\right] + \mathcal{O}[(b-b_{c})\ln(b-b_{c})].
\fe
Moreover, the regular contribution $I_R(b)$ is defined as
\ie
I_{R}(b) = \int^{0}_{1} f_{R}(z,b_{c})\mathrm{d}z + \mathcal{O}[(b-b_{c})\ln(b-b_{c})].
\fe
Let us define the regular part of the integrand as $f_R = f(z, r_0) - f_D(z, r_0)$. Under the strong deflection approximation, the total deflection angle then takes the form
\ie
\label{strongdeflecccc}
a(b) = - \Tilde{a} \ln \left[ \frac{b}{b_{c}}-1    \right] + \Tilde{b} + \mathcal{O}[(b-b_{c})\ln(b-b_{c})],
\fe
with
\ie
\Tilde{a} = \sqrt{\frac{2 \mathcal{B}_{ph}A_{ph}}{\mathcal{C}^{\prime\prime}_{ph}\mathcal{A}_{ph} - \mathcal{C}_{ph}\mathcal{A}^{\prime\prime}_{ph}}}, \,\,\,\,\,\,\,\, \text{and} \,\,\,\,\,\,\,\, \Tilde{b} = \Tilde{a} \ln\left[ r^{2}_{ph}\left( \frac{\mathcal{C}^{\prime\prime}}{\mathcal{C}_{ph}}  -  \frac{\mathcal{A}^{\prime\prime}_{ph}}{\mathcal{C}_{ph}} \right)   \right] + I_{R}(r_{ph}) - \pi.
\fe


\subsection{Charged black hole influenced by an antissymetric tensor}

Now, particularly, let us focus on the metric (\ref{metrictsuka}) \cite{duan2023electrically}:
\ie
\nonumber
\mathrm{d}s^{2} = - \left( \frac{1}{1-l} - \frac{2M}{r} + \frac{Q^{2}}{(1-l)^{2}r^{2}}    \right) \mathrm{d}t^{2} + \frac{\mathrm{d}r^{2}}{\left( \frac{1}{1-l} - \frac{2M}{r} + \frac{Q^{2}}{(1-l)^{2}r^{2}}    \right)} + r^{2} \mathrm{d} \theta^{2} + r^{2}\sin^{2}\theta \mathrm{d}\varphi^{2}.
\fe

In the asymptotic limit, the metric functions exhibit the following behavior: \(\lim\limits_{r \to \infty} A(r) = \frac{1}{1-l}\), \(\lim\limits_{r \to \infty} B(r) = 1-l\), and \(\lim\limits_{r \to \infty} C(r) = r^{2}\). At first glance, this appears to contradict the condition of asymptotic flatness imposed in the previous section. However, since \(l\) is a constant, one can rescale the time and radial coordinates as \( t \to t^{\prime} = \frac{t}{\sqrt{1-l}} \) and \( r \to r^{\prime} = \sqrt{1-l} \, r \). Given that \(l\) is small, representing the Lorentz violation coefficient, the metric in this limit simplifies accordingly
\begin{equation}
  \lim\limits_{r^{\prime} \to \infty}\mathrm{d}s^{2}=-\mathrm{d}t^{\prime 2}+\mathrm{d}r^{\prime 2}+\left(1-\frac{l}{2}+ \mathcal{O}(l^2)\right)r^{\prime 2}\left(\mathrm{d}\theta^2 +\sin^{2}{\theta}\mathrm{d}\phi^2\right).
  \label{reff}
\end{equation}

Notably, this line element bears resemblance to that of a global monopole with a deficit solid angle \(\delta = \frac{l}{2}\), accurate to first order in \(l\) \cite{PhysRevLett.63.341}. As discussed in the previous section, due to the spherical symmetry, photon trajectories can be confined to the equatorial plane (\(\theta = \pi/2\)). Under this condition, the line element simplifies to:
\begin{equation}
     \lim\limits_{r^{\prime} \to \infty}\mathrm{d}s^{2} |_{\theta=\pi/2}=-\mathrm{d}t^{\prime 2}+\mathrm{d}r^{\prime 2}+\left(1-\frac{l}{2}+ \mathcal{O}(l^2)\right)r^{\prime 2}\mathrm{d}\phi^2.
     \label{lin}
\end{equation}

This spacetime can be interpreted as a cosmic string geometry intersected by planes of constant \(z\). Consequently, the line element \eqref{lin} is locally flat with a conical topology, and it is inherently not asymptotically flat due to the presence of a deficit angle \(\delta = \frac{l}{2}\). At first glance, this might suggest that the general methodology outlined in the previous section is inapplicable. However, the requirement of asymptotically flat spacetimes is introduced primarily to ensure the positivity of the effective potential at infinity (\(r \to \infty\)), which guarantees the existence of photons at large distances (as discussed earlier and in \cite{tsukamoto2017deflection}). In this context, such a condition can be relaxed since \(\lim\limits_{r \to \infty} V(r) = E^2 > 0\) is satisfied.

In this section, it is pertinent to emphasize that we exclusively consider the solution without a cosmological constant. This restriction is due to the applicability of the Tsukamoto method \cite{tsukamoto2017deflection}. To the black hole under consideration, the corresponding photon sphere is written as \cite{heidari2024impact}
\ie
r_{ph} = \frac{\sqrt{9 (l-1)^4 M^2+8 (l-1) Q^2}+3 (l-1)^2 M}{2 (1-l)},
\fe
and, by considering $l$ small, we obtain
\ie
\label{photonsphere}
r^{(l\ll 0)}_{ph} \approx \frac{1}{2} \left(\sqrt{9 M^2-8 Q^2}+3 M\right) + \frac{ \left(-3 M \sqrt{9 M^2-8 Q^2}-9 M^2-4 Q^2\right)l}{2 \sqrt{9 M^2-8 Q^2}},
\fe
where it gives rise to a natural restriction of the parameter $Q$ and $M$ which is
\ie
9 M^2-8 Q^2 > 0, \quad \text{and} \quad 9 M^2-8 Q^2 \neq 0,
\fe
ensuring that the expression yields real and positive values. It is worth noting that the first term in Eq. (\ref{photonsphere}) coincides with the result for the Reissner–Nordström black hole. In contrast, the second term arises purely as a correction introduced by the Lorentz--violating parameter $l$.

Having outlined the methodology above, we proceed to apply it to our metric described in Eq. (\ref{metrictsuka}). Thus, we obtain 
\ie
b_{c} = 3 \sqrt{3-3 l} (1-l) M -\frac{\sqrt{3-3 l} \, Q^2}{2 \left((l-1)^2 M\right)}.
\fe
In addition, $\Tilde{a}$ and $\Tilde{b}$ can be written as 
\ie
\Tilde{a} = 1 + \frac{Q^2}{9 M^2}  +\left(\frac{5 Q^2}{18 M^2}-\frac{1}{2}\right)l.
\fe
Therefore, we have 
\ie
\begin{split}
& \Tilde{b} =  \left[ 1 + \frac{Q^2}{9 M^2} +  \left(\frac{5 Q^2}{18 M^2}-\frac{1}{2}\right) l \right] \left(\ln (6)-\frac{(3 l+1) Q^2}{9 M^2}\right)
+ I_{R}(r_{ph}) - \pi.
\end{split}
\fe

In contrast to the Schwarzschild and Reissner--Nordström cases, the contribution to the parameter $\Tilde{a}$ is significantly altered by the presence of Lorentz violation $l$, as expected. Also, $I_{R}(r_{ph})$ reads 
\ie
\begin{split}
  I_{R}(r_{ph}) =   \int_{0}^{1} \mathrm{d}z  & \left( - \frac{1}{18 M^2 (3-2 z)^{3/2} z} \right) \times  \left\{  -18 (l-2) M^2 (2 z-3) \left(\sqrt{3}-\sqrt{3-2 z}\right) \right. \\
 & \left. -(5 l+2) Q^2 \left(3 \sqrt{3} z^2-8 \sqrt{3} z+4 \sqrt{3-2 z} z-6 \sqrt{3-2 z}+6 \sqrt{3}\right)  \right\} \\
 & = \frac{(5 l+2) Q^2 \left(2 \sqrt{3}-5+\ln (36)+\ln \left(7-4 \sqrt{3}\right)\right)}{18 M^2}-2 (l-2) \ln \left(3-\sqrt{3}\right).
\end{split}
\fe
Consequently, the deflection angle present in Eq. (\ref{strongdeflecccc}) is expressed as 
\ie
\begin{split}
a(b) = &  - \left[ 1+ \frac{Q^2}{9 M^2}+  \left(\frac{5 Q^2}{18 M^2}-\frac{1}{2}\right) l \right] \left[ \frac{b}{3 \sqrt{3-3 l} \,(1-l) M -\frac{\sqrt{3-3 l} Q^2}{2 \left((l-1)^2 M\right)}  } -1 \right] \\
& + \left[1 + \frac{Q^2}{9 M^2}+\left(\frac{5 Q^2}{18 M^2}-\frac{1}{2}\right)l \right] \left(\ln (6)-\frac{(3 l+1) Q^2}{9 M^2}\right) \\
&  + \frac{(5 l+2) Q^2 \left(2 \sqrt{3}-5+\ln (36)+\ln \left(7-4 \sqrt{3}\right)\right)}{18 M^2}-2 (l-2) \ln \left(3-\sqrt{3}\right)    - \pi \\
& \mathcal{O} \left[ \left( b - 3 \sqrt{3-3 l} (1-l) M -\frac{\sqrt{3-3 l} \, Q^2}{2 \left((l-1)^2 M\right)}\right) \ln \left(b - 3 \sqrt{3-3 l} (1-l) M -\frac{\sqrt{3-3 l} \, Q^2}{2 \left((l-1)^2 M\right)} \right) \right]  .
\end{split}
\fe

\begin{figure}
    \centering
     \includegraphics[scale=0.51]{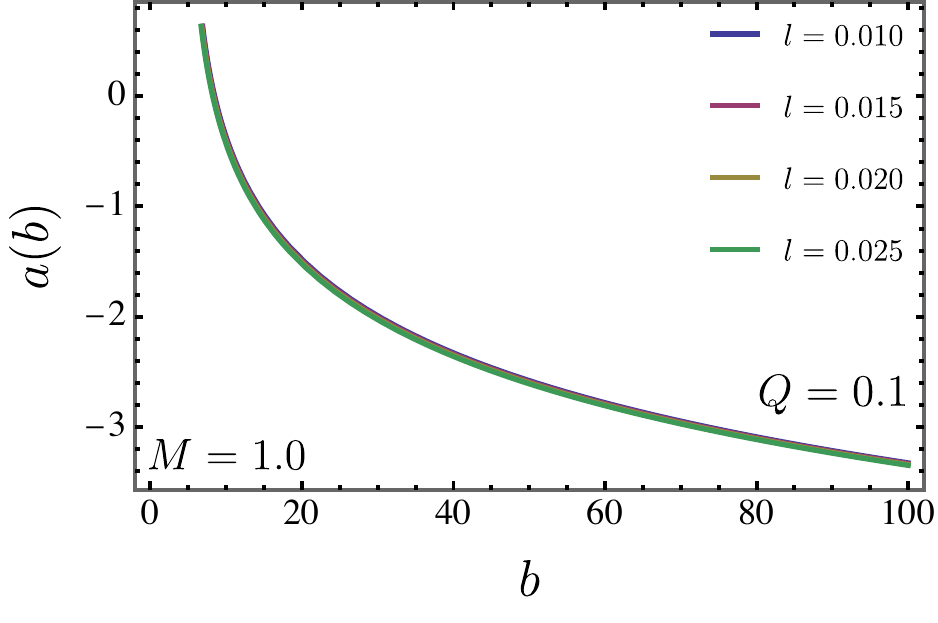}
    \includegraphics[scale=0.51]{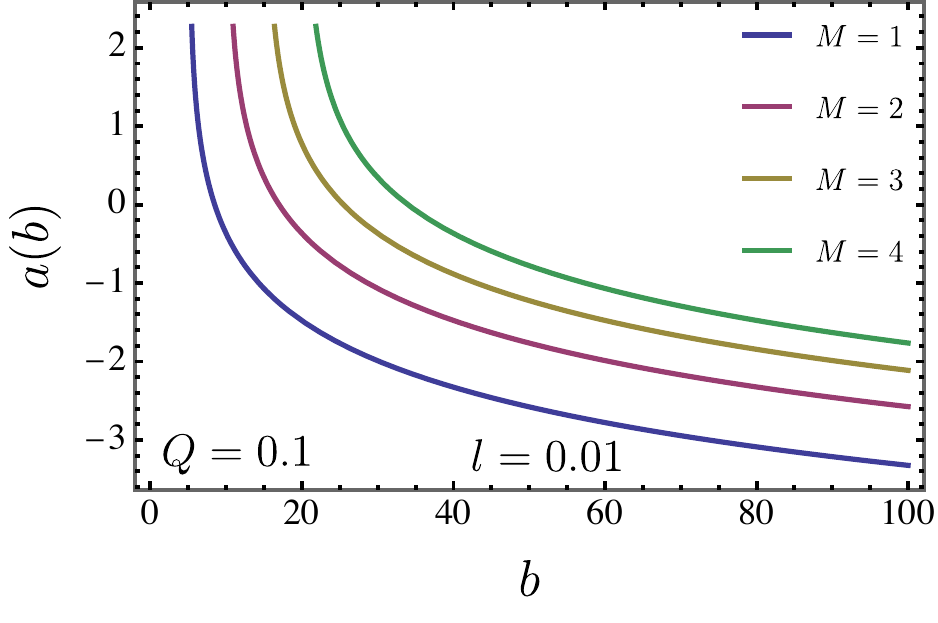}
     \includegraphics[scale=0.51]{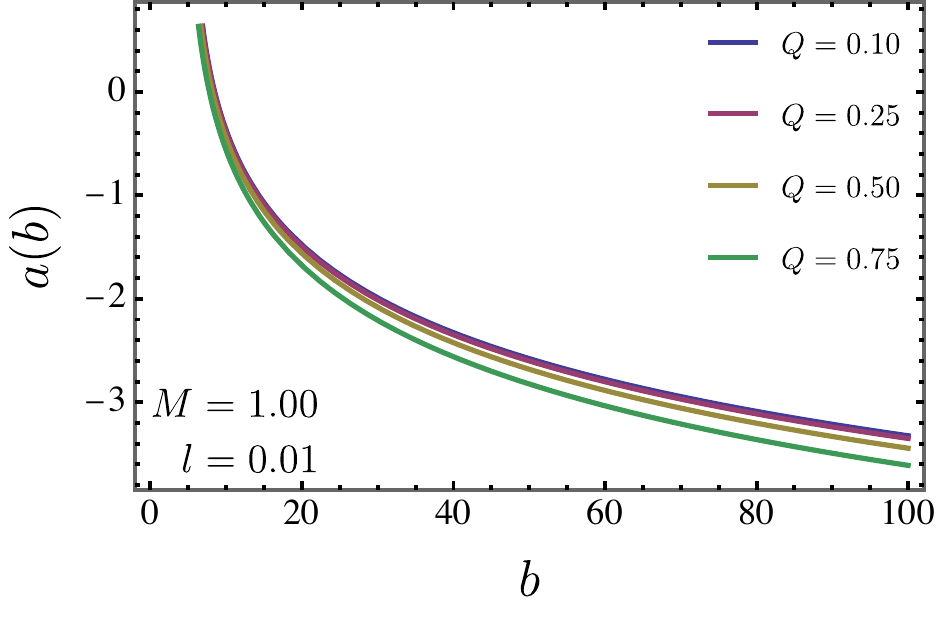}
    \caption{The deflection angle as a function of $b$ for different values of $l$, $M$ and $Q$.}
    \label{sdjidjsi}
\end{figure}

To facilitate understanding, we present in Fig. \ref{sdjidjsi} the variation of the deflection angle with $b$ under various system conditions. In the top--left panel, although the variations are subtle, an increase in the parameter $l$ results in a slight decrease in the magnitude of the deflection angle $a(b)$, considering $Q = 0.1$ and $M = 1$. In the top--right panel, keeping $Q = 0.1$ and $l = 0.01$ fixed, a larger mass $M$ leads to an increase in the magnitude of $a(b)$. Finally, the bottom panel reveals that raising the charge $Q$, while maintaining $M = 1$ and $l = 0.01$, causes a reduction in the deflection angle.

\section{Lensing equations and observables}

\begin{figure}
       	\centering
       	\begin{tikzpicture}[scale=0.9]
       	\node (I)    at ( 5,-0.75)   {$L$};
         \node (I)    at ( 6.8,0.9)   {$\theta$};
       	\node (II)    at ( 10,-0.5)   {$O$};
       	\node (II)    at ( -0.5,1.5)   {$S$};
       	\node (II)    at ( -0.5,5)   {$I$};
       	\node    at ( 2.5,-1.5)   {$D_{LS}$};
       	\node    at ( 7.5,-1.5)   {$D_{OL}$};
       	
       	\draw (10,0)--(0,0)--(0,5);
       	\draw [thick,rounded corners=20pt] (0,1.5)--(5,2.5)--(10,0);
       	\draw [dashed] (5,2.5)--(0,5);
       	\draw [dashed](10,0)--(0,1.5);
        \draw [thick,fill](5,0) circle (16pt);
       	\fill[orange] (5,0) circle (2pt);
        \fill[purple] (10,0) circle (2pt);
       	\fill[red] (0,1.5) circle (2pt);
       	\fill[blue] (0,5) circle (2pt);
       	\draw
       	(6,2) coordinate (a) 
       	-- (5,2.5) coordinate (b) 
       	-- (6,2.7) coordinate (c) 
       	pic["$a$", draw=red, <->, angle eccentricity=1.2, angle radius=0.9cm]
       	{angle=a--b--c};
       	
       	\draw
       	(4,2.3) coordinate (d) 
       	-- (5,2.5) coordinate (e) 
       	-- (4,3) coordinate (f) 
       	pic["$a$", draw=red, <->, angle eccentricity=1.2, angle radius=0.9cm]
       	{angle=f--e--d};
       	
       	\draw
       	(8,0) coordinate (g) 
       	-- (10,0) coordinate (h) 
       	-- (8,1) coordinate (i) 
       	pic[ draw=black, angle eccentricity=1.05, <->, angle radius=2.8cm]
       	{angle=i--h--g};
       	
       	\draw
       	(7,0) coordinate (g1) 
       	-- (10,0) coordinate (h1) 
       	-- (7,0.47) coordinate (i1) 
       	pic["$\beta$", draw=black, angle eccentricity=1.1, <->, angle radius=2.35cm]
       	{angle=i1--h1--g1};	
       	
       	\draw[ <->,
       	decoration={markings,
       		mark= at position 0.5 with {\arrow{|}},
       	},
       	postaction={decorate}
       	]
       	(0,-1) node[anchor=west] {} -- (10,-1);
       	
       	\end{tikzpicture}
       	\caption{Representation of the gravitational lensing. The light emitted from the source \( S \) (red point) is bent as it travels toward the observer \( O \) (purple point), influenced by the presence of a compact object positioned at \( L \) (orange point). The observer \( O \) perceives an image \( I \) (blue point). \( D_{OL} \) represents the distance between the lens \( L \) and the observer \( O \), while \( D_{LS} \) denotes the distance from the source's projection to the lens along the optical axis. The big black dot represents the black hole under consideration }
       	\label{asLrG}
       \end{figure}
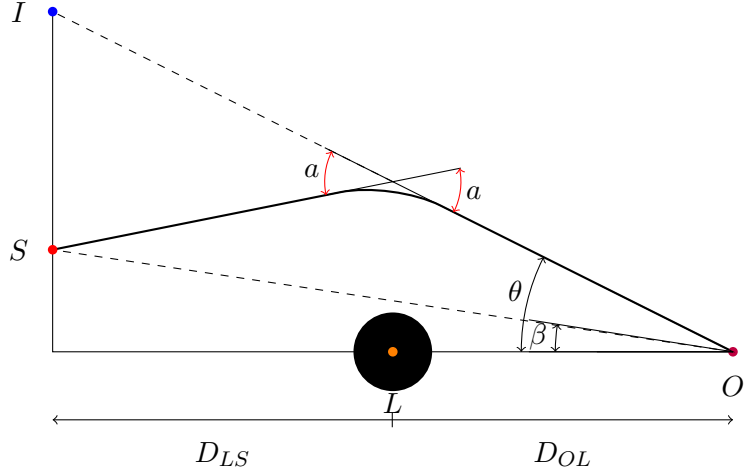

In this section, various parameters related to the bending of light in the strong gravitational field of the black hole are investigated. Fig. \ref{asLrG} demonstrates the gravitational lensing effect caused by the black hole described by Eq. \eqref{metrictsuka}. Notice that light emitted from the source \( S \) (red point) is deflected towards the observer \( O \) (purple point) due to the gravitational influence of the Lorentz--violating black hole at point \( L \) (orange point). The image seen by the observer is represented by \( I \) (blue point). The big black dot accounts for the black hole studied here. The angular positions of the source and the observed image are labeled as \(\beta\) and \(\theta\), respectively. The angular deviation of the light, \( a \), shows the change in its path as it travels through the gravitational field.

Additionally, we use the same setup proposed in \cite{030,bozza2001strong}, where the source (\( S \)) is nearly perfectly aligned with the lens (\( L \)). This scenario is significant because it produces relativistic images. Under these conditions, the lens equation describing the relationship between \(\theta\) and \(\beta\) is given by
$\beta=\theta-\frac{D_{LS}}{D_{OS}}\Delta a_{n}$.
It is important to mention that \(\Delta a_{n}\) is defined as the deflection angle after accounting for all the loops completed by the photons before reaching the observer, specifically given by \(\Delta a_{n} = a - 2n\pi\). In this approach, the impact parameter is approximated as \(\Tilde{b} \approx \theta D_{OL}\). Therefore, the angular deviation is expressed as $
\label{angle}
a(\theta)=- \Tilde{a}\ln\left(\frac{\theta D_{OL}}{b_c}-1\right)+\Tilde{b}$.

To derive \(\Delta a_{n}\), we expand \( a(\theta) \) around \( \theta = \theta^{0}_n \), satisfying the condition \( \alpha(\theta^{0}_n) = 2n\pi \) $
\Delta a_{n}=\frac{\partial a}{\partial\theta}\Bigg|_{\theta=\theta^0_n}(\theta-\theta^0_n) \ $.
We have $
\theta^0_{n}=\frac{b_c}{D_{OL}}\left(1+e_n\right), \qquad\text{where}\quad e_n=e^{\Tilde{b}-2n\pi} \ $. Also, equation \( \Delta a_{n}=-\frac{\Tilde{a}D_{OL}}{b_ce_n}(\theta-\theta^0_n) \) can be obtained. This result is then integrated into the lens equation, leading to the derivation of the expression for the \( n^{th} \) angular position of the image $
\theta_n\simeq\theta^0_n+\frac{b_ce_n}{\Tilde{a}}\frac{D_{OS}}{D_{OL}D_{LS}}(\beta-\theta^0_n) \ $. The deflection of light preserves surface brightness, while the gravitational lens modifies the solid angle of the source, affecting its observable appearance. The total flux received from a relativistic image depends on the magnification \( \mu_{n} \), defined as \( \mu_n=\left|\frac{\beta}{\theta}\frac{\partial\beta}{\partial\theta}\bigg|_{\theta^0_{n}}\right|^{-1} \). By regarding \( \Delta a_{n}=-\frac{\Tilde{a}D_{OL}}{b_ce_n}(\theta-\theta^0_n) \), we derive $
\mu_{n}=\frac{e_n(1+e_n)}{\Tilde{a}\beta}\frac{D_{OS}}{D_{LS}}\left(\frac{b_c}{D_{OL}}\right)^2 \ . $

Notice that the magnification factor \( \mu_n \) rises as \( n \) increases, indicating that the brightness from the initial image \( \theta_1 \) far exceeds that of subsequent images. Nevertheless, the overall luminosity remains diminished, primarily due to the presence of the term \( \left(\frac{b_c}{D_{OL}}\right)^2 \). An important observation is the divergence in magnification as \( \beta \to 0 \), underscoring that optimal alignment between the lens and the source maximizes the potential for detecting relativistic images, as one should expect. Furthermore, it is worthy to be mentioned that the impact parameter may effectively be correlated to \( \theta_{\infty} \), as elaborated in \cite{bozza2001strong}
$ b_c=D_{OL}\theta_{\infty} \ , $
where $\theta_{\infty}$ denotes the other relativistic images here.
We will adopt Bozza's approach from \cite{bozza2001strong}, which treats the outermost image \(\theta_{1}\) as a distinct entity, while grouping the remaining images under \(\theta_{\infty}\). To elaborate, Bozza introduced the following observables $
       s=\theta_{1}-\theta_{\infty}= \theta_{\infty} e^{\frac{\Tilde{b}-2\pi}{\Tilde{a}}} \ ,
       \tilde{r} =  \frac{\mu_{1}}{\sum\limits_{n=2}^{\infty} \mu_{n} }= e^{\frac{2\pi}{\Tilde{a}}} \ $.

In the expressions mentioned earlier, \( s \) indicates the angular separation, while \( \tilde{r} \) denotes the ratio of the flux emitted by the first image to the total flux emitted by all other images combined. These equations can be inverted to find the expansion coefficients. To substantiate our findings, the following subsection will demonstrate a particular astrophysical example to compute these observables and investigate the influence of the Lorentz--violating parameter \( l \) under these previous quantities.

\subsection{Galactic phenomena: gravitational lensing by Sagittarius $A^{*}$}

Observational data on stellar dynamics compellingly indicates the presence of a dense, mysterious entity at the center of our galaxy. This entity, believed to be the supermassive black hole Sagittarius (Sgr) $A^{*}$, has an estimated mass of \(4.4 \times 10^6 M_{\odot}\) \cite{genzel2010galactic}. To better understand this celestial phenomenon, we examine its characteristics using the dimensionless parameter \(l\), which helps elucidate the behavior of observables.

\begin{figure}
    \centering
     \includegraphics[scale=0.41]{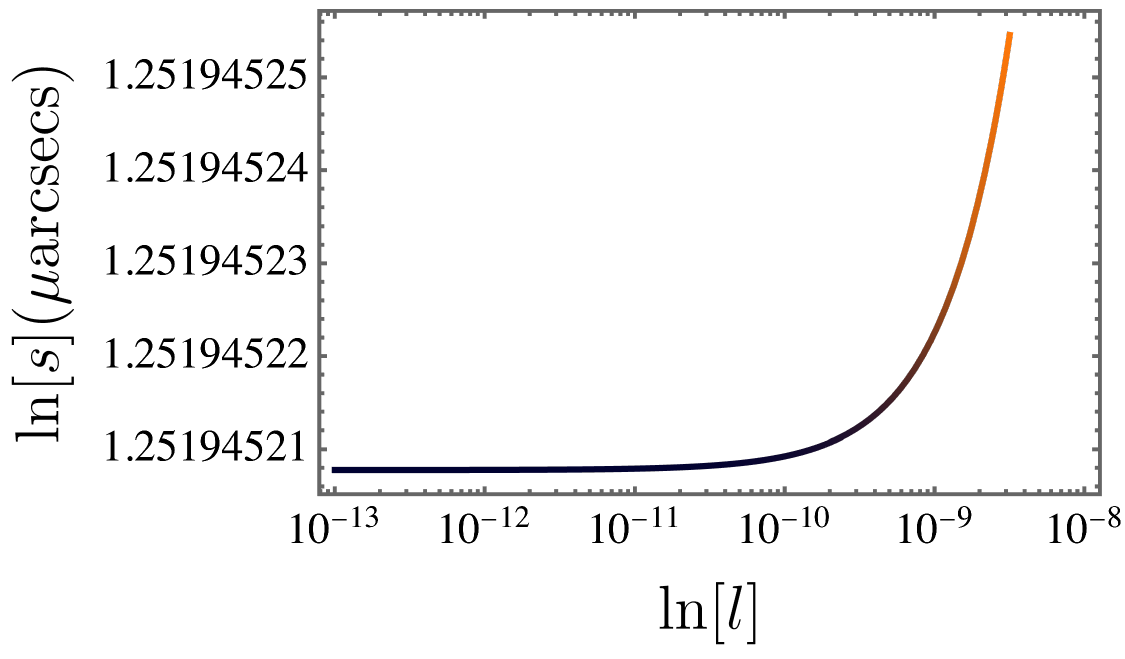}
    \caption{Observable $\ln s$ for distinct values of $\ln l$ and for $Q=0.1$.}
    \label{obs1}
\end{figure}

\begin{figure}
    \centering
     \includegraphics[scale=0.4]{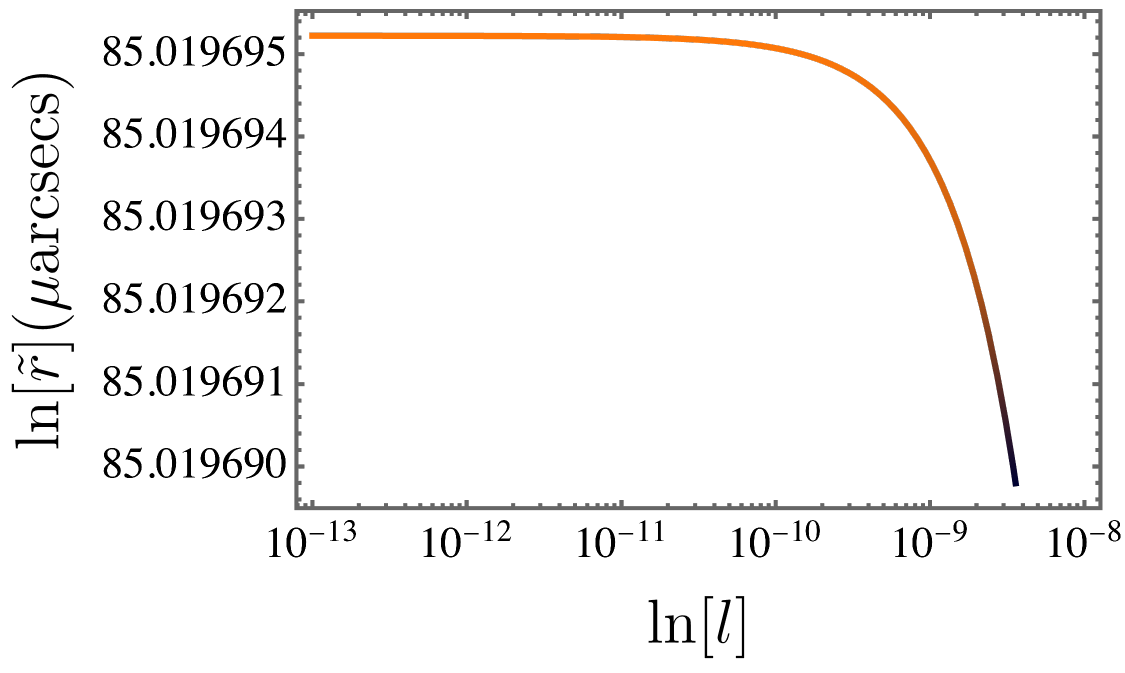}
    \caption{Observable $\ln \Tilde{r}$ for a variety of $\ln l$ and for $Q=0.1$.}
    \label{obs2}
\end{figure}

To analyze the observables, we adopt a distance of $D_{OL} = 8.5$ Kpc \cite{genzel2010galactic} and \( l \sim 3.90708 \times 10^{-12} \) based on literature \cite{junior2024spontaneous}. With $ b_{c} =  3 \sqrt{3-3 l} (1-l) M -\frac{\sqrt{3-3 l} \, Q^2}{2 \left((l-1)^2 M\right)}$, we find \( \theta_{\infty} \approx 25.64 \) $\mu$arcsecs \( + \mathcal{O}(l) \), where \( \mathcal{O}(l) \) denotes first--order effects in the Lorentz violation parameter, approximately \( 10^{-12} \). For a better interpretation, Figs. \ref{obs1} and \ref{obs2} illustrate observables \( s \) and \( \Tilde{r} \) with respect to the LV coefficient. Both figures employ logarithmic scales due to the small contribution of LV. Fig. \ref{obs1} reveals that \( s \) increases as \( l \) rises, indicating greater separation between the first and subsequent relativistic images. Conversely, Fig. \ref{obs2} demonstrates that the flux ratio of the first image decreases with increasing \( l \).

In addition, the gravitational lensing was also addressed in the context of charged Simpsson--Visser solution \cite{zhang2022gravitational}. As argued in this reference, the angular size \( \theta_{\infty} \) for Sgr $A^{*}$ ranges from approximately $20.7$ to $26.6$ $\mu$as, with deviations \( \delta \theta_{\infty} \) between $-6.0$ and $0$ $\mu$as. Both \( \theta_{\infty} \) and \( \delta \theta_{\infty} \) decrease as the charge \( q \) increases \cite{zhang2022gravitational}. While \( \theta_{\infty} \) is within the detection limits of the Event Horizon Telescope (EHT), the small deviation \( \delta \theta_{\infty} \) of up to 6 $\mu$as is currently beyond its detection capability. Therefore, it is not possible to differentiate between the black--bounce--Reissner--Nordström spacetime and the Schwarzschild black hole based on \( \delta \theta_{\infty} \).


\section{Time delay analysis: theoretical framework}

This section introduces the theoretical framework of our study. The time delay of light in a gravitational field is calculated by solving the differential equations of null geodesics in a spherically symmetric spacetime
$\mathrm{d}\tau^{2} = f(r)\mathrm{d}t^{2} -\frac{1}{f(r)}\mathrm{d}r^{2} 
	-r^{2}(\mathrm{d}\theta^{2}+\sin^{2}\theta \mathrm{d}\phi^{2})$
so that the conserved quantities give rise to $L \equiv  r^{2}\sin^{2}\theta \frac{\mathrm{d}\phi}{\mathrm{d}\lambda}$, $E  \equiv   f(r)\frac{\mathrm{d}t}{\mathrm{d}\lambda}$, and $\mathcal{L}  \equiv  g_{\mu\nu}\mathrm{d}x^{\mu}\mathrm{d}x^{\nu} 
		= f(r) \bigg( \frac{\mathrm{d}t}{\mathrm{d}\lambda} \bigg)^2
		- \frac{1}{f(r)} \bigg( \frac{\mathrm{d}r}{\mathrm{d}\lambda} \bigg)^{2}
		- r^{2} \bigg( \frac{\mathrm{d}\theta}{\mathrm{d}\lambda} \bigg)^{2} 
		- r^{2}\sin^{2}\theta \bigg( \frac{\mathrm{d}\phi}{\mathrm{d}\lambda} \bigg)^{2}$,
where $\lambda$ is an affine parameter, $L$ denotes the conserved angular momentum, and $E^2/2$ represents the conserved energy along a particle's orbit. For test particles confined to the equatorial plane ($\theta = \pi/2$), these ones lead to the following simplified differential equations
\ie
		\frac{1}{2} \bigg( \frac{\mathrm{d}r}{\mathrm{d}\lambda} \bigg)^{2} + \frac{1}{2} f(r) \bigg[ \frac{L^{2}}{r^{2}} + \mathcal{L} \bigg]
		= \frac{1}{2} \bigg( \frac{\mathrm{d}r}{\mathrm{d}\lambda} \bigg)^{2} + V(r)
		= \frac{1}{2}E^{2}.
\fe
In this context, \( V(r) = \frac{f(r)}{2} \left[ \frac{L^2}{r^2} + \mathcal{L} \right] \) defines the effective potential for particles in a spherically symmetric gravitational field, with the impact parameter given by \( b \equiv |L/E| \). For massless particles on null geodesics, \( \mathcal{L} = 0 \). Focusing on photon trajectories, we obtain:
\ie
	\frac{\mathrm{d}r}{\mathrm{d}t} = \frac{\mathrm{d}r}{\mathrm{d}\lambda}  \frac{\mathrm{d}\lambda}{\mathrm{d}t}
	= \pm f(r) \sqrt{1 - b^{2}\frac{f(r)}{r^{2}}}  .
\fe
Using \( E = f(r) \frac{\mathrm{d}t}{\mathrm{d}\lambda} \) and setting \( \mathcal{L}=0 \) for a massless photon, the signs \( \pm \) can be interpreted as follows. As a particle moves along its scattering orbit from the source position \( r_{\text{S}} \), the radial coordinate \( r \) decreases over time until the particle reaches the closest approach, \( r = r_0 \), to the central black hole. Beyond this turning point \( r = r_0 \), the radial coordinate begins to increase as time advances. Thus, we establish the following relationships:
\ie
\frac{\mathrm{d}r}{\mathrm{d}t} = - f(r) \sqrt{1 - b^{2} \frac{f(r)}{r^{2}}} < 0 ,
\fe
for a photon traveling from the initial position \( r = r_{\text{S}} \) to the turning point \( r = r_{0} \), the radial coordinate behavior is described. Additionally,
\ie
\frac{\mathrm{d}r}{\mathrm{d}t} = f(r) \sqrt{1 - b^{2} \frac{f(r)}{r^{2}}} > 0,  
\fe
where we have considered that the photon moves from the tuning point $r=r_{0}$ to observer position $r=r_{\text{O}}$. It is worth commenting that, in gravitational lensing, when the light source is situated at \( r = r_{\text{S}} \) and the observer at \( r = r_{\text{O}} \), the time delay of light as it propagates through the gravitational field is formulated as \cite{qiao2024time}
\ie
\begin{split}
 \Delta T & = T - T_{0} \\
 & = -\int_{r_{\text{S}}}^{r_{0}} \frac{\mathrm{d}r}{f(r)\sqrt{1-\frac{b^{2} f(r)}{r^{2}}}}
	      + \int_{r_{0}}^{r_{\text{O}}} \frac{\mathrm{d}r}{f(r)\sqrt{1-\frac{b^{2} f(r)}{r^{2}}}}
	      - T_{0}
	      \\
	& = \int_{r_{0}}^{r_{\text{S}}} \frac{\mathrm{d}r}{f(r)\sqrt{1-\frac{b^{2} f(r)}{r^{2}}}}
          + \int_{r_{0}}^{r_{\text{O}}} \frac{\mathrm{d}r}{f(r)\sqrt{1-\frac{b^{2} f(r)}{r^{2}}}}
          - \sqrt{r_{\text{S}}^{2}-r_{0}^{2}} - \sqrt{r_{\text{O}}^{2}-r_{0}^{2}}, 
\end{split}
\fe
where $T_{0} = \sqrt{r_{\text{S}}^{2} - r_{0}^{2}} + \sqrt{r_{\text{O}}^{2} - r_{0}^{2}} $ represents the time taken for light to propagate in the absence of a gravitational field.
If we assume that the parameters $Q$, $l$, and the impact parameter are small, the time delay can be expressed analytically as follows
\ie
\begin{split}
\Delta T  = & \, 2 (l-1) r_{0} -\frac{8 l M^2}{r_{0} - 2 M} + \frac{4 l M Q^2}{(r_{0}-2 M)^2}-\frac{2 Q^2}{r_{0} - 2 M} + (1-l)r_{\text{O}} -\frac{2 l M Q^2}{(r_{\text{O}}-2 M)^2}\\
& +\frac{4 l M^2}{r_{\text{O}} -2 M}+\frac{Q^2}{r_{\text{O}} - 2 M} + (1-l)r_{\text{S}} + \frac{4 l M^2}{r_{\text{S}} - 2 M}-\frac{2 l M Q^2}{(r_{\text{S}} - 2 M)^2} +\frac{Q^2}{r_{\text{S}} - 2 M} \\
& +b^2 \left(\frac{1}{r_{0}}-\frac{r_{\text{O}} + r_{\text{S}}}{2 r_{\text{O}} r_{\text{S}}}\right) + 2 (2l - 1) M \Big[ 2 \ln (r_{0} - 2 M)  -\ln (r_{\text{O}} - 2 M) - \ln (r_{\text{S}} - 2 M) \Big] \\
& - \sqrt{r_{\text{S}}^{2} - r_{0}^{2}} - \sqrt{r_{\text{O}}^{2} - r_{0}^{2}}.
\end{split}
\fe

\begin{figure}
    \centering
     \includegraphics[scale=0.51]{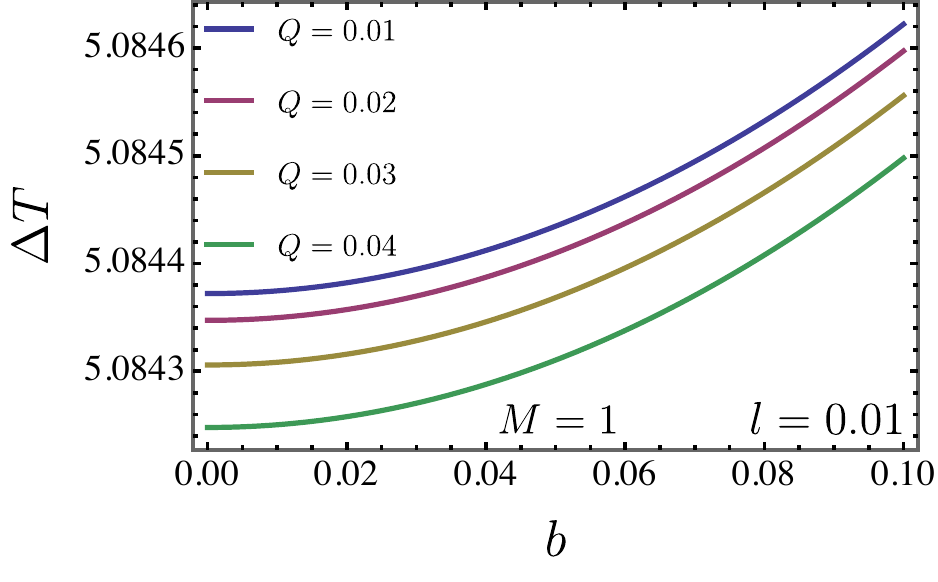}
     \includegraphics[scale=0.48]{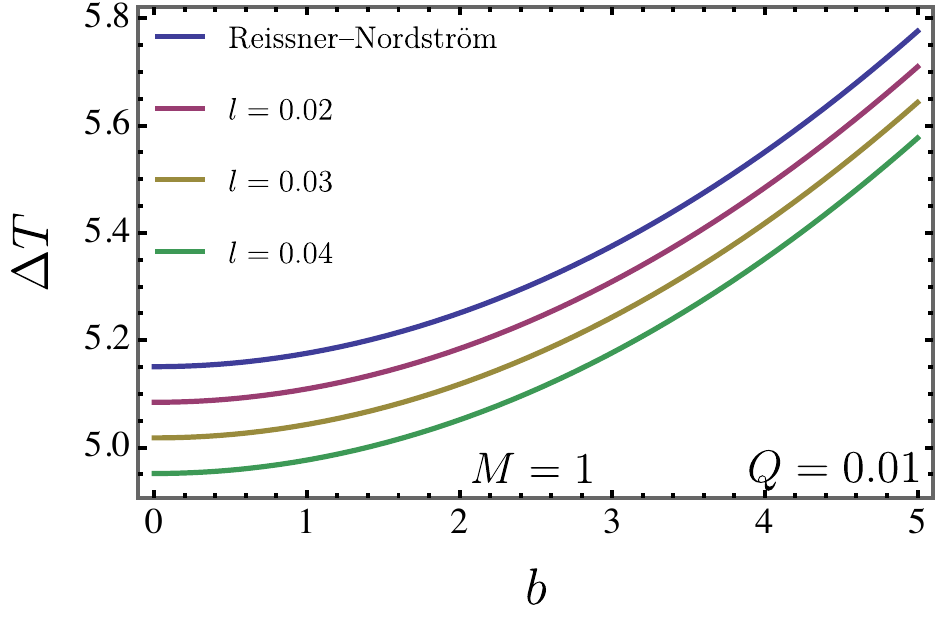}
    \caption{The time delay $\Delta T$ is plotted as a function of the impact parameter $b$, with the left panel illustrating the behavior for various values of the charge $Q$, while the right panel displays the effect of the Lorentz--violating parameter $l$.}
    \label{timedelay}
\end{figure}

\begin{figure}
    \centering
       \includegraphics[scale=0.46]{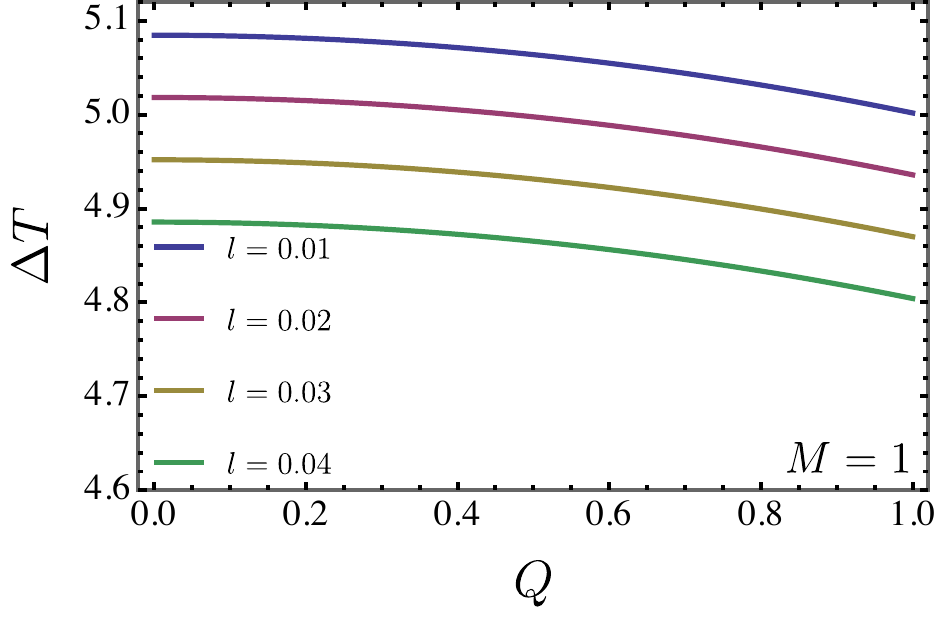}
     \includegraphics[scale=0.52]{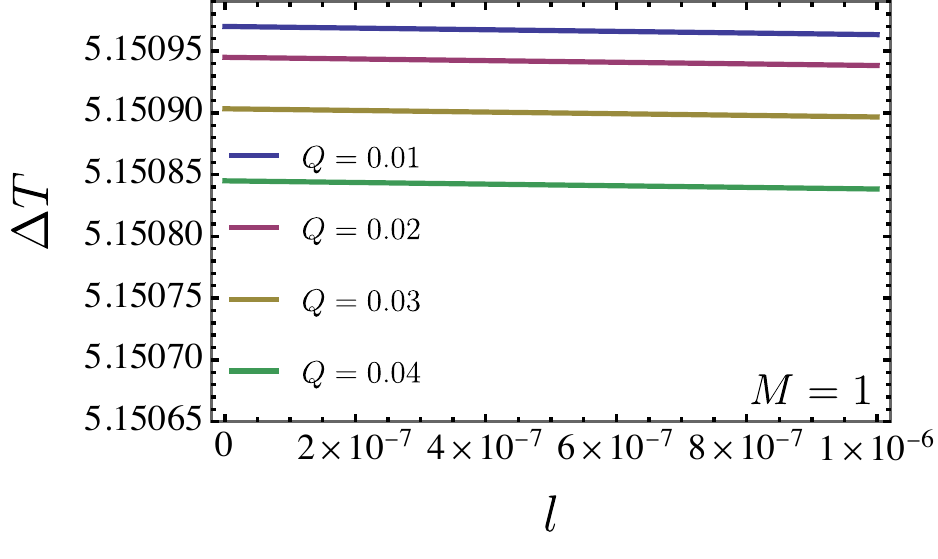}
    \caption{The time delay $\Delta T$ is presented as a function of the charge $Q$ for different values of the Lorentz--violating parameter $l$ in the left panel, while the right panel displays $\Delta T$ as a function of $l$ for several fixed values of $Q$.}
    \label{timedelay2}
\end{figure}

Figs. \ref{timedelay} and \ref{timedelay2} illustrate the behavior of the time delay $\Delta T$ as a function of the impact parameter $b$ and the parameters $Q$ and $l$, respectively. In Fig. \ref{timedelay}, the left panel displays $\Delta T$ versus $b$ for various values of $Q$, indicating that an increase in the charge parameter $Q$ leads to a noticeable decrease in the magnitude of the time delay. Conversely, the right panel reveals that increasing the Lorentz--violating parameter $l$ also results in a reduction of $\Delta T$.

In Fig. \ref{timedelay2}, we further analyze the combined influence of $Q$ and $l$ on $\Delta T$. The left panel examines the variation of $\Delta T$ with $Q$ for different values of $l$, while the right panel explores the dependence on $l$ for fixed values of $Q$. Overall, the results suggest that increasing $Q$ tends to suppress the time delay across a broad range of $l$ values, whereas variations in $l$ yield only a slight reduction in $\Delta T$ for different values of $Q$.

\section{Summary and conclusion}
\label{summary}

This study investigated the gravitational lensing phenomena associated with a spherically symmetric charged black hole in the presence of Lorentz symmetry violation, modeled through an antisymmetric Kalb--Ramond tensor field.

The analysis was conducted within both the weak and strong deflection regimes. In the weak deflection limit, we employed the \textit{Gauss--Bonnet} theorem to derive analytical expressions for the deflection angle $\alpha(b, l, Q)$. The results showed that, for increasing values of the Lorentz--violating parameter $l$ and the electric charge $Q$, the deflection angle decreased as a function of the impact parameter $b$. However, when the black hole mass $M$ was increased while keeping $l$ and $Q$ fixed, the magnitude of $\alpha(b, l, Q)$ increased.

In the strong deflection limit, we adopted the \textit{Tsukamoto} method, which allowed us to compute key observables such as the positions and magnifications of relativistic images. Notably, these results were also derived analytically. We applied this framework to the supermassive black hole Sagittarius $A^{*}$ and evaluated the corresponding observables as functions of the Lorentz--violating parameter $l$. Within this approach, as $a(b)$ varied, both the charge $Q$ and the parameter $l$ contributed to a reduction in the deflection angle when the mass $M$ was fixed. In contrast, for fixed values of $Q$ and $l$, an increase in $M$ led to an enhancement of $a(b)$.

Lastly, we also computed the time delay. In general, as $b$ increased, the magnitude of $\Delta T$ decreased with increasing $Q$, for fixed $M$ and $l$. A similar trend was observed for increasing $l$ at fixed values of $M$ and $Q$.

\section*{Acknowledgments}
\hspace{0.5cm} A. A. Araújo Filho is supported by Conselho Nacional de Desenvolvimento Cient\'{\i}fico e Tecnol\'{o}gico (CNPq) and Fundação de Apoio à Pesquisa do Estado da Paraíba (FAPESQ), project No. 150891/2023-7. 
Special thanks are extended to Ali Övgün and P. J. Porfírio for the insightful correspondence throughout the preparation and revision of the manuscript. The author also wishes to thank N. Heidari for highlighting a specific reference that was fundamental in the time delay calculation.

\section{Data Availability Statement}

Data Availability Statement: No Data associated in the manuscript

\bibliographystyle{ieeetr}
\bibliography{main}

\end{document}